\documentclass[pre,aps,amsmath,amssymb,amsfonts,floatfix,superscriptaddress,showpacs,twocolumn,footinbib]{revtex4-1}
\usepackage{graphicx}
\usepackage{amsmath,amsfonts}
\usepackage{xcolor}
\usepackage{color}
\usepackage{mathtools}
\usepackage{eufrak}
\usepackage{pgfplots}
\usepackage{soul}
\usepackage{multirow}
\usepackage{comment}

\newcommand{\up}{\uparrow}
\newcommand{\dn}{\downarrow}
\newcommand{\kv}{\ensuremath{\mathbf{k}}}

\newcommand{\qv}{\ensuremath{\mathbf{q}}}

\newcommand{\ch}{\ensuremath{\text{ch}}}
\newcommand{\sz}{\ensuremath{\text{sp}}}

\newcommand{\pp}{\ensuremath{{pp}}}

\newcommand{\trip}{\ensuremath{\text{t}}}
\newcommand{\sing}{\ensuremath{\text{s}}}
\newcommand{\firr}{\ensuremath{\text{Uirr}}}

\usepackage{tikz}

\usetikzlibrary{calc}
\usetikzlibrary{decorations.pathmorphing}
\usetikzlibrary{decorations.pathreplacing}
\usetikzlibrary{decorations.markings}
\usetikzlibrary{shapes.misc}
\usetikzlibrary{shapes}
\usetikzlibrary{positioning}
\usetikzlibrary{snakes}
\usetikzlibrary{arrows}
\usetikzlibrary{fadings}
\usetikzlibrary{calc,arrows}
\tikzstyle{decision} = [diamond, draw, fill=blue!20, text width=4.5em, text badly centered, node distance=3cm, inner sep=0pt]
\tikzstyle{block} = [rectangle, draw, fill=blue!20, text width=7em, text centered, rounded corners, minimum height=5em]
\tikzstyle{line} = [draw, -latex']
\tikzstyle{cloud} = [draw, ellipse,fill=red!20, node distance=3cm, minimum height=2em]
\tikzstyle{overbrace style}=[decorate,decoration={brace,raise=2mm,amplitude=3pt}]
\tikzstyle{overbrace text style}=[font=\footnotesize, above, pos=.5, yshift=3mm]
\tikzset{snake it/.style={decorate, decoration=snake}}
\usetikzlibrary{3d}
\usepackage{mathtools}
    \tikzset{
            partial ellipse/.style args={#1:#2:#3}{
                        insert path={+ (#1:#3) arc (#1:#2:#3)}
                            }
                        }

\usetikzlibrary{calc}
\tikzset{
            inertial frame/.style = {x={(-20:2cm)}, y={(-160:2cm)}, z={(90:2cm)}},
              local frame/.style = {shift={(local origin)}, x={(40:.7cm)}, y={(150:.7cm)}, z={(105:.7cm)}}
          }

    \tikzset{middlearrow/.style={
                decoration={markings,
                            mark= at position 0.65 with {\arrow{#1}} ,
                                    },
                                            postaction={decorate}
                                                }
                                                }
\tikzset{cross/.style={cross out, draw, 
         minimum size=2*(#1-\pgflinewidth), 
                  inner sep=0pt, outer sep=0pt}}

\def\presuper#1#2%
  {\mathop{}%
   \mathopen{\vphantom{#2}}^{#1}%
   \kern-0.5\scriptspace%
   #2}

\usepackage{hyperref}
\hypersetup{colorlinks=true,breaklinks,linkcolor=blue,urlcolor=blue,citecolor=blue}
\begin{document}

    \pgfmathdeclarefunction{gauss}{2}{%
          \pgfmathparse{1/(#2*sqrt(2*pi))*exp(-((x-#1)^2)/(2*#2^2))}%
          }
    \pgfmathdeclarefunction{mgauss}{2}{%
          \pgfmathparse{-1/(#2*sqrt(2*pi))*exp(-((x-#1)^2)/(2*#2^2))}%
          }
    \pgfmathdeclarefunction{lorentzian}{2}{%
        \pgfmathparse{1/(#2*pi)*((#2)^2)/((x-#1)^2+(#2)^2)}%
          }
    \pgfmathdeclarefunction{mlorentzian}{2}{%
        \pgfmathparse{-1/(#2*pi)*((#2)^2)/((x-#1)^2+(#2)^2)}%
          }

\author{Friedrich Krien}
\email{krien@ifp.tuwien.ac.at}
\affiliation{Jo\v{z}ef Stefan Institute, Jamova 39, SI-1000, Ljubljana, Slovenia}
\affiliation{Institute for Solid State Physics, TU Wien, 1040 Vienna, Austria}

\author{Alexander I. Lichtenstein}
\affiliation{Institute of Theoretical Physics, University of Hamburg, 20355 Hamburg, Germany}

\author{Georg Rohringer}
\affiliation{Institute of Theoretical Physics, University of Hamburg, 20355 Hamburg, Germany}


\title{Fluctuation diagnostic of the nodal/antinodal dichotomy in the Hubbard model at weak coupling:
A parquet dual fermion approach}

\begin{abstract}
We apply the boson exchange parquet solver for dual fermions to the
half-filled Hubbard model on a square lattice at small interaction.
Our results establish that, in this regime, nonlocal vertex corrections
play an important role in the formation of the pseudogap.
Namely, in comparison to the simpler ladder approximation,
these additional vertex corrections included in the parquet equations enhance
the coupling of spin fluctuations with the quasiparticles.
The pseudogap thus opens already at a higher temperature,
in quantitative agreement with the numerically exact diagrammatic Monte Carlo.
The representation of the parquet diagrams in terms of
boson exchange facilitates large lattice sizes and gives rise to an unbiased fluctuation diagnostic
of the self-energy, which does not rely on the Fierz ambiguity.
The fluctuation diagnostic implies that nodal and antinodal fermions are affected 
equally by spin fluctuations with the exact commensurate nesting vector $(\pi,\pi)$.
However, the antinode couples more efficiently to incommensurate
fluctuations than the node, leading to the nodal/antinodal dichotomy.
We corroborate this finding in terms of a spin-fermion-like calculation.

\end{abstract}

\maketitle

\section{Introduction}
The half-filled Hubbard model on the simple square lattice exhibits strong
antiferromagnetic spin fluctuations at low temperature,
which lead to the opening of a spectral gap for arbitrarily small
interaction~\cite{Vilk96,Vilk97,Schaefer15,Schaefer16,vanLoon18-2,Schaefer20}.
This can be seen as an emergent strong-coupling regime where,
in spite of a weak Hubbard repulsion much smaller than the bandwidth,
a strong effective interaction between electrons is mediated by spin fluctuations.
The corresponding spin fluctuation theory and the phenomenological spin-fermion model
capture indeed qualitatively the gap formation~\cite{Vilk97,Katanin09,Schaefer20}.

A crucial feature of the temperature-driven crossover from metal to insulator
is the {\sl nodal/antinodal dichotomy}:
Upon lowering the temperature, the gap opens first at the antinode,
then spreads across the Fermi surface, until also the node becomes
insulating~\cite{Schaefer15,Schaefer16,vanLoon18-2,Schaefer20}.

In the considered regime, thanks to recent improvements of
the diagrammatic Monte Carlo (DiagMC) method~\cite{Prokofev98,Rossi16,Kozik19},
the gap formation at weak Hubbard interaction was recently confirmed
in a numerically exact framework~\cite{Schaefer20}.
However, it remains interesting to apply approximate techniques to the problem,
which allow us either to simplify the picture,
and thus identify the minimal ingredients that describe the correct physics,
or to obtain a more complete picture by gathering additional information
about the fermionic and bosonic excitations and their interaction.
With regard to the first route, the authors of Ref.~\cite{Schaefer20} compared the DiagMC to the 
spin fluctuation theory and to the ladder dynamical vertex approximation~\cite{Toschi07,Katanin09},
which describe qualitatively the gap formation and give access
to quasiparticle parameters~\cite{Vilk97,Rohringer16}.

In this work, we proceed along the second route in the spirit of the  {\sl fluctuation diagnostic} 
method for the self-energy~\cite{Gunnarsson15,Gunnarsson16,Gunnarsson18,Arzhang20} which provides 
substantial information about the feedback of collective bosonic fluctuations on the fermions. 
We use this idea to determine the origin of the nodal/antinodal dichotomy of the self-energy. 
We arrive at the conclusion that {\sl in}commensurate spin fluctuations precipitate the dichotomy, 
whereas commensurate spin fluctuations with the exact nesting vector $(\pi,\pi)$ 
lead to the same feedback on nodal and antinodal fermions. 



The authors of Ref.~\cite{Schaefer20} also provide a detailed comparison of the DiagMC with a variety of approximate techniques,
some of which capture qualitatively the pseudogap formation, for example,
the two-particle self-consistent approach~\cite{Vilk97},
the ladder dynamical vertex approximation,
and the ladder dual fermion approach (LDFA,~\cite{Rubtsov08,Hafermann09}).
On the other hand, the quantitative deviations of these methods from DiagMC with respect
to crossover temperatures and/or magnitude of the self-energy are sizable.

In fact, it is a common conception that spin fluctuations are dominant at half-filling and that, hence,
the ladder approximations should be sufficient because the interplay between various fluctuation channels can be neglected.
However, this argument is not very conclusive
because in a more complete theory of vertex corrections
the strong spin fluctuations renormalize not only other one- and two-particle correlations, but also themselves,
via their feedback on the kernel of the Bethe-Salpeter equation.
This feedback is neglected in the ladder approximations and
we show here that the more complete parquet summation of
vertex diagrams~\cite{Dominicis64,Dominicis64-2}
applied to dual fermions~\cite{Astretsov19,Astleithner20}
improves the quantitative agreement with DiagMC.
The good agreement with the numerically exact reference result indicates that all relevant vertex corrections, also beyond the ladder approximation, are taken into account in our approach.
Indeed, we observe a nodal/antinodal dichotomy of the fermion-spinboson coupling,
an effect which can not be captured by the ladder approximation.

The methodological improvements which make the parquet formalism applicable to reasonably large lattices (in this work up to $32\times32$ lattice sites) are described
in Refs.~\cite{Astretsov19,Eckhardt20,Krien20,Krien20-2,Wentzell20},
where the accompanying paper~\cite{Krien20} presents an efficient
{\sl boson exchange parquet solver (BEPS)} for dual fermions, which we use here.
The method can be construed as a partial bosonization with residual four-fermion interaction~\cite{Denz19}, 
combined with a truncated unity approximation~\cite{Lichtenstein17,Eckhardt20}
that allows a feasible solution of a set of parquet equations for the residual four-fermion vertex~\cite{Krien19-2,Krien19-3,Krien20}.
Similar to the dynamical mean-field theory (DMFT,~\cite{Georges96}),
the dual fermion approach relies on an auxiliary Anderson impurity model (AIM)
as a nontrivial starting point that provides the local correlations~\cite{Rubtsov08}.
We find that in the pseudogap regime the self-consistent adjustment of the
hybridization bath of the AIM achieves an optimal agreement with DiagMC.
Finally, as an addendum to Ref.~\cite{Krien20},
we show here explicitly the decomposition of the BEPS self-energy into contributions
corresponding to single- and multi-boson exchange~\cite{Krien19-2}.
This decomposition implies directly a fluctuation diagnostic of the self-energy
which is unambiguous, that is, it does not make use of the Fierz ambiguity to rewrite the self-energy in a charge, spin, or particle-particle picture~\cite{Gunnarsson15}.

The paper is structured as follows. In Sec.~\ref{sec:model} we introduce the Hubbard model
and the key aspects of our method, which is described in full detail in Ref.~\cite{Krien20}.
In Sec.~\ref{sec:diagnostic} we define the fluctuation diagnostic.
The results are presented in Sec.~\ref{sec:results}, we close with the conclusions in Sec.~\ref{sec:conclusions}.

\section{Model and Method}\label{sec:model}
We consider the paramagnetic Hubbard model,
\begin{align}
    H = &-t\sum_{\langle ij\rangle\sigma} c^\dagger_{i\sigma}c^{}_{j\sigma}+ U\sum_{i} n_{i\up} n_{i\dn},\label{eq:hubbardmodel}
\end{align}
on the square lattice, where ${t}$ is the hopping between nearest-neighbor lattice sites $i,j$ which we use as energy unit (i.e., ${t}=1$).
$c^{},c^\dagger$ are the annihilation and creation operators, $\sigma=\up,\dn$ is the spin index,
which is suppressed where unambiguous.
$U$ is the Hubbard repulsion between two electrons at the same site.
We consider here the weak coupling regime, $U/t=2$.

\subsection{Anderson impurity model}\label{sec:aim}
The dual fermion approach, which we will use to solve the model in Eq.~(\ref{eq:hubbardmodel}), is based on an auxiliary AIM with the imaginary time action,
\begin{align}
  S_{\text{AIM}}=&-\sum_{\nu\sigma}c^*_{\nu\sigma}(\imath\nu+\mu-h_\nu)c^{}_{\nu\sigma}+U \sum_\omega n_{\up\omega} n_{\dn\omega},
  \label{eq:aim}
\end{align}
where $c^*,c$ are Grassmann numbers and $\nu$ and $\omega$ 
are fermionic and bosonic Matsubara frequencies, respectively.
Summations over Matsubara frequencies $\nu, \omega$ contain implicitly the factor $T=\beta^{-1}$,
the temperature. We consider two different options to fix the hybridization function $h_\nu$
of the AIM, as we will discuss in Sec.~\ref{sec:sc}.
To calculate the correlation functions of the AIM~\eqref{eq:aim} at particle-hole symmetry
we employ a continuous-time quantum Monte-Carlo (CTQMC) solver~\cite{Bauer2011}
with improved estimators~\cite{Hafermann12}.
We require several higher (three- and four-point) correlation functions of the AIM,
and a decomposition of the four-point vertex introduced in Ref.~\cite{Krien19-2},
called single-boson exchange (SBE) decomposition.
The definitions correspond, precisely, to the ones of Sec.~IIB of Ref.~\cite{Krien20}.

\subsection{Dual fermions}
In the dual fermion formalism~\cite{Rubtsov08} the Hubbard model~\eqref{eq:hubbardmodel} is mapped to the dual action,
\begin{align}
S[d^*,d]=&-\sum_{k\sigma}G^{0,-1}_kd^*_{k\sigma}d_{k\sigma}\label{eq:dualaction}\\
+&\frac{1}{4}\sum_{kk'q}\sum_{\sigma_i}f^{\sigma_1\sigma_2\sigma_3\sigma_4}_{\nu\nu'\omega}
d^*_{k\sigma_1}d^*_{k'+q,\sigma_2}d_{k'\sigma_3}d_{k+q,\sigma_4}.\notag
\end{align}

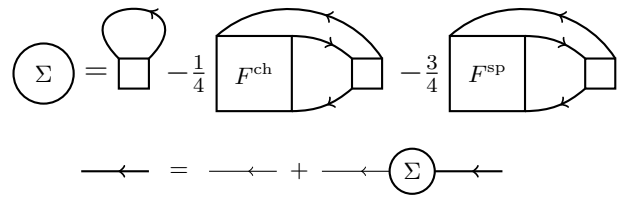
\begin{figure}
\begin{center}
\begin{tikzpicture}
\begin{scope}[shift={(-2.8,0)}]
        \draw[thick] (0.5,0) circle [radius=0.4cm];
        \draw[thick] (.5,0) node{$\Sigma$};
        \draw[thick] (1.2,0) node{\large$=$};
        \draw[thick,middlearrow={<}] (1.5,.2) to [out=135,in=45,looseness=8] (1.9,.2);
        \draw[thick] (1.5,-.2) -- (1.5,.2) -- (1.9,.2) -- (1.9,-.2) -- cycle;
        
\end{scope}
        \draw[thick] (-.4,0) node{\large$-\frac{1}{4}$};
        \draw[thick] (0,-.5) -- (0,.5) -- (1.,.5) -- (1.,-.5) -- cycle;
        \draw[thick] (.5,0) node{${F}^\ch$};
        \draw[thick,middlearrow={>}] (1.,.5) to [bend left=20] (1.8,.2);
	    \draw[thick,middlearrow={<}] (1.,-.5) to [bend right=20] (1.8,-.2);
	    \draw[thick] (1.8,-.2) -- (1.8,.2) -- (2.2,.2) -- (2.2,-.2) -- cycle;
	    \draw[thick,middlearrow={<}] (0.,.5) to [bend left=50] (2.2,.2);

\begin{scope}[shift={(3.1,0)}]
        \draw[thick] (-.4,0) node{\large$-\frac{3}{4}$};
        \draw[thick] (0,-.5) -- (0,.5) -- (1.,.5) -- (1.,-.5) -- cycle;
        \draw[thick] (.5,0) node{${F}^\sz$};
        \draw[thick,middlearrow={>}] (1.,.5) to [bend left=20] (1.8,.2);
	    \draw[thick,middlearrow={<}] (1.,-.5) to [bend right=20] (1.8,-.2);
	    \draw[thick] (1.8,-.2) -- (1.8,.2) -- (2.2,.2) -- (2.2,-.2) -- cycle;
	    \draw[thick,middlearrow={<}] (0.,.5) to [bend left=50] (2.2,.2);
\end{scope}
      \begin{scope}[shift={(-2,-1.3)}]
        \begin{scope}[shift={(.2,0)}]
        \draw [thick,middlearrow={<}] (0,0) -- (0.9,0);
        \end{scope}
        \draw (1.5,0) node{$=$};
        \begin{scope}[shift={(.5,0)}]
        \draw [middlearrow={<}] (1.4,0) -- (2.3,0);
        \draw (2.6,0) node{$+$};
        \draw [middlearrow={<}] (2.9,0) -- (3.8,0);
        \draw[thick] (4.1,0) circle [radius=0.3cm];
        \draw[thick] (4.1,0) node{$\Sigma$};
        \draw [thick,middlearrow={<}] (4.4,0) -- (5.3,0);
        \draw (6,0) node{};
      \end{scope}
      \end{scope}
\end{tikzpicture}
\end{center}
    \caption{\label{fig:sigma}
    Top: Dual self-energy. Arrows denote the dual Green's function $G$,
    large boxes represent the vertex function $F$ in parquet approximation, small boxes the impurity vertex $f$.
    Bottom: Dyson equation, thin arrows represent the bare dual Green's function $G^0$. 
    }
    \end{figure}

Here, $k=(\kv,\nu)$ and $q=(\qv,\omega)$ denote fermionic and bosonic momentum-energy vectors, respectively.
The Grassmann numbers $d^*,d$ represent the dual fermions and the
dual bare propagator is the nonlocal DMFT Green's function, $G^0=G^\text{DMFT}-g$, where
\begin{align}
G^\text{DMFT}_k=\frac{1}{\imath\nu-\varepsilon_\kv+\mu-\Sigma^\text{imp}_\nu}.\label{eq:dmft}
\end{align}
Here, $\varepsilon_\kv$ denotes the dispersion of the square lattice.
The local Green's function $g_\sigma(\nu)=-\langle c^{}_{\nu\sigma}c^*_{\nu\sigma}\rangle$ and the
local self-energy $\Sigma^\text{imp}_\nu$ are obtained from the AIM~\eqref{eq:aim}.

The dual action in Eq.~(\ref{eq:dualaction}) contains, in principle, also three- and more-particle local interactions.
Following the standard dual fermion applications, we keep only the quartic interaction between the dual fermions,
given by the four-point vertex $f_{\nu\nu'\omega}$ of the AIM~\eqref{eq:aim}.

In any approximation beyond DMFT the bare propagator $G^0$ is dressed with a dual self-energy,
\begin{align}
G_k=\frac{G^0_k}{1-G^0_k\Sigma_k}.\label{eq:dyson}
\end{align}
The self-energy reads in the general case (see Fig.~\ref{fig:sigma}),
\begin{align}
{\Sigma}_k=&\;\Sigma^{\text{HF}}_\nu\label{eq:sigma}\\
-&\frac{1}{4}\sum_{k'q}{G}_{k+q}\left[{F}^\ch_{kk'q}{X}^0_{k'q}f^\ch_{\nu'\nu\omega}
+3{F}^\sz_{kk'q}{X}^0_{k'q}f^\sz_{\nu'\nu\omega}\right].\notag
\end{align}
Here, $X^0_{kq}=G_kG_{k+q}$ denotes the dual bubble and $F$ is the full vertex function
in the parquet approximation~\footnote{
The parquet approximation for dual fermions implies that the vertex $\tilde{\Lambda}$
which is fully irreducible with respect to pairs of (dual) Green's function lines
is approximated by the bare dual interaction, $\tilde{\Lambda}_{kk'q}\approx f_{\nu\nu'\omega}$.
In principle, the diagrammatic content of this approximation is comparable to the parquet
dynamical vertex approximation~\cite{Toschi07,Eckhardt20},
but the dual fermions are not plagued by divergences of the irreducible vertex~\cite{Schaefer13}.}
for dual fermions~\cite{Astretsov19}.
The Hartree-Fock part is $\Sigma^{\text{HF}}_\nu=\sum_{k'}{G}_{k'}f^\ch_{\nu'\nu,\omega=0}$.
Labels ``$\ch, \sz$'' and, further below, ``$\sing, \trip$''
denote the charge, spin, particle-particle singlet, and particle-particle triplet channel, respectively.
Summations over momenta imply division by the number of lattice sites
$N$~\footnote{$\sum_k$ implies a factor $(\beta N)^{-1}$.}.

We use the BEPS method described in Ref.~\cite{Krien20} to obtain
a self-consistent solution for the dual vertex function $F$ and self-energy $\Sigma$.
The {\sl lattice} self-energy of the Hubbard model~\eqref{eq:hubbardmodel}
is given as~\cite{Rubtsov08},
\begin{align}
\Sigma^\text{lat}_k=\Sigma^\text{imp}_\nu+\frac{\Sigma_k}{1+g_\nu\Sigma_k}.\label{eq:sigma_lat}
\end{align}

\subsection{Self-consistency conditions}\label{sec:sc}
We consider two different choices for the hybridization bath $h_\nu$ of the AIM in Eq.~\eqref{eq:aim}.
The \textit{first} corresponds to the DMFT solution~\cite{Georges96} for the Hubbard model~\eqref{eq:hubbardmodel},
where the local part of the DMFT lattice Green's function is adjusted to the Green's function $g$ of the AIM,
\begin{align}
G^\text{DMFT}_{ii}(\nu)=g(\nu).\label{eq:sc_dmft}
\end{align}

The \textit{second} choice for the hybridization corresponds to an outer self-consistency loop of the dual fermion method.
The hybridization $h_\nu$ is then fixed so that the local part of the dual fermion propagator vanishes,
\begin{align}
G_{ii}(\nu)=0.\label{eq:sc_df}
\end{align}
The effect of this prescription is that any diagram with a local dual line vanishes.
This prescription is superior to others~\cite{Ribic18,vanLoon18-3},
as it indeed improves LDFA results in comparison with numerically exact benchmarks~\cite{Gukelberger17}.
One should note that, when using Eq.~\eqref{eq:sc_df}, the Green's function $G^\text{DMFT}$
in Eq.~\eqref{eq:dmft} is no longer equal to the DMFT solution. However,
we keep the label ``DMFT'' also for this case, since the functional form is not changed by the outer self-consistency
and the bare dual Green's function is still given as $G^0=G^\text{DMFT}-g$.

\section{Fluctuation diagnostic}\label{sec:diagnostic}
In this work, we employ the feasible reformulation of the parquet approximation
for dual fermions introduced in Ref.~\cite{Krien20}, called a \textit{boson exchange parquet solver} (BEPS).
The reader should keep in mind that this method employs a {\sl double} decomposition of the full vertex,
first, with respect to single-boson exchange (SBE,~\cite{Krien19-2}),
and second, in the sense of the parquet equations~\cite{Bickers04,Kauch19}.
We show in the following that the two decompositions give rise to a fine-structured fluctuation diagnostic
of the self-energy in terms of single- and multi-boson exchange.

\begin{figure}[b]
\begin{center}
\begin{tikzpicture}
        \begin{scope}[shift={(-1.6,0)}]
        \draw[thick] (0,-.5) -- (0,.5) -- (.8,0) -- cycle;
        \draw [decorate,decoration={snake,amplitude=1pt, segment length=9pt},very thick] (0.8,0) -- (1.8,0) ;
        \draw[thick] (1.8,0) -- (2.6,-.5) -- (2.6,.5) -- cycle;
        \end{scope}
        \draw[thick,middlearrow={>}] (1.,.5) to [bend left=20] (1.8,.2);
	    \draw[thick,middlearrow={<}] (1.,-.5) to [bend right=20] (1.8,-.2);
	    \draw[thick] (1.8,-.2) -- (1.8,.2) -- (2.2,.2) -- (2.2,-.2) -- cycle;
	    \draw[thick,middlearrow={<}] (-1.6,.5) to [bend left=50] (2.2,.2);
\begin{scope}[shift={(3.1,0)}]
        \begin{scope}[shift={(.5,-1)},rotate=90,scale=0.665]
        \begin{scope}[shift={(0,0)}]
        \draw[thick] (0,1.2) -- (1,1.2) -- (.5,.4) -- cycle;
        \draw [decorate,decoration={snake,amplitude=1pt, segment length=9pt},very thick] (.5,.4) -- (.5,-.4) ;
        \draw[thick] (0,-1.2) -- (1,-1.2) -- (.5,-.4) -- cycle;
        \end{scope}
        \begin{scope}[shift={(2.,0)}]
	    \draw[thick,middlearrow={<}] (-1,1.2) -- (0,1.2);
	    \draw[thick,middlearrow={>}] (-1,-1.2) -- (0,-1.2);
        \draw[thick] (0,1.2) -- (1,1.2) -- (.5,.4) -- cycle;
        \draw [decorate,decoration={snake,amplitude=1pt, segment length=9pt},very thick] (.5,.4) -- (.5,-.4) ;
        \draw[thick] (0,-1.2) -- (1,-1.2) -- (.5,-.4) -- cycle;
        \end{scope}
        \end{scope}
        \begin{scope}[shift={(.3,0)}]
        \draw[thick,middlearrow={>}] (1.,1) to [bend left=20] (1.8,.2);
	    \draw[thick,middlearrow={<}] (1.,-1) to [bend right=20] (1.8,-.2);
	    \draw[thick] (1.8,-.2) -- (1.8,.2) -- (2.2,.2) -- (2.2,-.2) -- cycle;
	    \draw[thick,middlearrow={<}] (-.6,1) to [bend left=70] (2.2,.2);
	    \end{scope}
\end{scope}
\end{tikzpicture}
\end{center}
    \caption{\label{fig:twoboson} Schematic single- and two-boson exchange diagrams for the self-energy.
    At half-filling and weak coupling the direct contribution of multi-boson exchange
    to the self-energy is negligible, however, through the parquet equations
    the corresponding vertex corrections enter also the fermion-boson coupling (triangles)
    and the screened interaction (wiggly lines), leading to a sizable renormalization of the left diagram.
    }
    \end{figure}
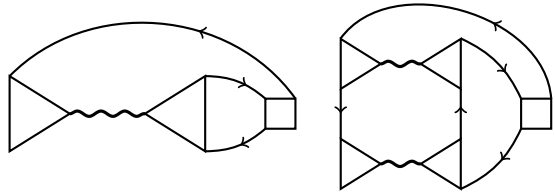

\subsection{Vertex decomposition}
In Ref.~\cite{Krien20} the vertex $F$ was expressed through single-boson exchange $\Delta$
and a residual vertex $\Phi^{\firr}$,
\begin{align}
    F^\alpha_{kk'q}\!=\!\Phi^{\firr,\alpha}_{kk'q}\!+\!\Delta^{ph,\alpha}_{kk'q}
    \!+\!\Delta^{\overline{ph},\alpha}_{kk'q}\!+\!\Delta^{\pp,\alpha}_{kk',q+k+k'}
    \!-\!2U^\alpha,\!\!\label{eq:jib_simple}
\end{align}
where $\alpha=\ch, \sz,\text{s}$ and $U^\ch=U, U^\sz=-U, U^\sing=2U$ is the bare interaction.
The SBE vertices are given as,
\begin{subequations}
\begin{align}
    \Delta^{ph,\alpha}_{kk'q}=&\Lambda^\alpha(k,q)W^\alpha(q)\Lambda^\alpha(k',q),\label{eq:nabla_simple}\\
    \Delta^{pp,\alpha}_{kk'q}=&\Lambda^{\sing\,}(k,q)W^{\sing\,}(q)\,\Lambda^{\sing\,}(k',q)
    \frac{1-2\delta_{\alpha,\sz}}{2},\label{eq:nabla_simple_pp}
\end{align}
\end{subequations}
where $W$ denotes the screened interaction and $\Lambda$ is the fermion-boson coupling
(also called the Hedin or proper vertex~\cite{Giuliani05,Sadovskii19}),
for definitions see Refs.~\cite{Krien19-2,Krien20}.

Further, the authors of Ref.~\cite{Krien20} introduced a second decomposition,
a parquet decomposition of the residual vertex $\Phi^\firr$,
\begin{widetext}
\begin{subequations}
\begin{align}
{\Phi}^{\firr,\ch}_{kk'q}=&\varphi^{\firr,\ch}_{\nu\nu'\omega}+{M}^{ph,\ch}_{kk'q}
-\frac{1}{2}{M}^{ph,\ch}_{k,k+q,k'-k}-\frac{3}{2}{M}^{ph,\sz}_{k,k+q,k'-k}+\frac{1}{2}{M}^{pp,\sing}_{kk',k+k'+q}
\;+\frac{3}{2}{M}^{pp,\trip}_{kk',k+k'+q}\label{eq:parquet_ch},\\
{\Phi}^{\firr,\sz}_{kk'q}=&\varphi^{\firr,\sz}_{\nu\nu'\omega}+{M}^{ph,\sz}_{kk'q}
-\frac{1}{2}{M}^{ph,\ch}_{k,k+q,k'-k}+\frac{1}{2}{M}^{ph,\sz}_{k,k+q,k'-k}-\frac{1}{2}{M}^{pp,\sing}_{kk',k+k'+q}
\;+\frac{1}{2}{M}^{pp,\trip}_{kk',k+k'+q},\label{eq:parquet_sp}
\end{align}
\end{subequations}
\end{widetext}
where $\varphi^\firr$ denotes the (local) residual vertex of the AIM~\eqref{eq:aim}\cite{Krien19-2,Krien19-3}.
Equations~\eqref{eq:parquet_ch} and \eqref{eq:parquet_sp}
are equivalent to the parquet approximation for dual fermions~\cite{Astretsov19}.

The double decomposition certainly implies a complication, however,
it amounts to the separation of high from low energies~\cite{Wentzell20},
as well as of short-ranged from long-ranged fluctuations~\cite{Krien20,Krien20-2}.
In particular, 
the spatially long-ranged components of the full vertex
$F$ are given by the SBE vertices $\Delta$~\cite{Krien20-2},
which also capture the high-frequency asymptote of $F$,
whereas multi-boson exchange represented by the vertices $M$
is more short-ranged compared to $\Delta$ and decays at high frequencies.
The short-ranged property of the vertices ${\Phi}^{\firr}$ and $M$
invites a truncated unity approximation~\cite{Eckhardt20} at this level.
For a detailed description of the various vertices and of the truncated unity approximation
for $\Phi^\firr$ the reader is referred to Ref.~\cite{Krien20}.

\subsection{Self-energy decomposition}\label{sec:fdiag}
In the BEPS method, vertex corrections are represented in terms of bosonic fluctuations.
This leads to an appealing physical picture where correlations can be assigned to various channels,
leading to a fluctuation diagnostic similar to Refs.~\cite{Gunnarsson15,Gunnarsson16}.
Using the SBE decomposition~\eqref{eq:jib_simple} we express the dual self-energy~\eqref{eq:sigma} as,
\begin{align}
\Sigma_k=\Sigma^{\text{HF}}_\nu+\Sigma^{\firr}_k+\Sigma^\ch_k+\Sigma^\sz_k
+\Sigma^\sing_k+{\Sigma}^{\text{bare}}_k.\label{eq:fdiag_sbe}
\end{align}
In a diagnostic of the lattice self-energy in Eq.~\eqref{eq:sigma_lat}
it is convenient to decompose only the numerator of the term $\Sigma/(1+g\Sigma)$,
because each component is anyways divided by the same denominator,
see Ref.~\cite{Arzhang20} and Sec.~\ref{sec:fdiag}.

Equation~\eqref{eq:fdiag_sbe} does not rely on the {\sl Fierz ambiguity}, that is,
the components $\Sigma^\ch, \Sigma^\sz, \Sigma^\sing$ can be assigned to one and only one fluctuation channel.
More precisely, the use of a renormalized/screened interaction can {\sl reduce} the Fierz ambiguity~\cite{Denz19}.
The dual fermions reduce the Fierz ambiguity as much as possible,
because their bare interaction corresponds to the full and,
hence, renormalized vertex function $f$ of the AIM~\eqref{eq:aim}.
However, there always remains some freedom in the choice of a channel for the bare Hubbard interaction $U$,
which plays a role also for the dual fermions as the leading order of the impurity vertex $f$.
Henceforth, we treat the bare interaction as a separate component, ${\Sigma}^{\text{bare}}$.

The last four components in Eq.~\eqref{eq:fdiag_sbe} then read,
\begin{subequations}
\begin{align}
{\Sigma}^{\ch}_k\!=\!-&\frac{1}{2}\!\sum_{k'q}\!{G}_{k+q}({\Delta}^{ph,\ch}_{kk'q}\!-\!U^\ch)
G_{k'}G_{k'+q}f^\ch_{\nu'\nu\omega}\label{eq:sbe_ch},\\
{\Sigma}^{\sz}_k\!=\!-&\frac{3}{2}\!\sum_{k'q}\!{G}_{k+q}({\Delta}^{ph,\sz}_{kk'q}\!-\!U^\sz)
G_{k'}G_{k'+q}f^\sz_{\nu'\nu\omega}\label{eq:sbe_sp},\\
{\Sigma}^{\sing}_k\!=\!-&\frac{1}{4}\!\sum_{k'{q}}\!{G}_{{q}-k}
({\Delta}^{pp,\sing}_{kk'{q}}-U^\sing)G_{k'}G_{q-k'}f^\sing_{\nu'\nu{\omega}},\label{eq:sbe_si}\\
{\Sigma}^{\text{bare}}_k\!=\!-&\frac{1}{4}\!\sum_{k'q}\!{G}_{k+q}UG_{k'}G_{k'+q}
\left[f^\ch_{\nu'\nu\omega}-3f^\sz_{\nu'\nu\omega}\right]\!.
\end{align}
\end{subequations}
In this way, the bare interaction is canceled from each vertex $\Delta$
and arises only in the term in the last line~\footnote{In other words,
we used the simple relation
$\Delta^{ph}+\Delta^{\overline{ph}}+\Delta^{\pp}-2U=(\Delta^{ph}-U)+(\Delta^{\overline{ph}}-U)+(\Delta^{\pp}-U)+U$.}. A self-energy diagram due to single-boson exchange is shown on the left of Fig.~\ref{fig:twoboson}.

We apply the \textit{second} vertex decomposition,
the parquet Eqs.~\eqref{eq:parquet_ch} and~\eqref{eq:parquet_sp} for the residual vertex $\Phi^\firr$,
which give rise to the following self-energy components:
\begin{align}
\Sigma^{\firr}_k=\Sigma^{\firr,\text{loc}}_k\!+\!\Sigma^{\firr,\ch}_k\!+\!\Sigma^{\firr,\sz}_k
\!+\!\Sigma^{\firr,\sing}_k\!+\!\Sigma^{\firr,\trip}_k,
\label{eq:fdiag_firr}
\end{align}
where the component $\Sigma^{\firr,\text{loc}}$ denotes the contribution
of the local residual vertex $\varphi^\firr_{\nu\nu'\omega}$,
the first term on the right-hand-sides of Eqs.~\eqref{eq:parquet_ch} and~\eqref{eq:parquet_sp}.
The other components are given explicitly as,
\begin{subequations}
\begin{align}
{\Sigma}^{\firr,\ch}_k=-&\frac{1}{2}\sum_{k'q}{G}_{k+q}
M^{ph,\ch}_{kk'q}G_{k'}G_{k'+q}f^\ch_{\nu'\nu\omega},\\
{\Sigma}^{\firr,\sz}_k=-&\frac{3}{2}\sum_{k'q}{G}_{k+q}
M^{ph,\sz}_{kk'q}G_{k'}G_{k'+q}f^\sz_{\nu'\nu\omega},\\
{\Sigma}^{\firr,\sing}_k=-&\frac{1}{4}\sum_{k'{q}}{G}_{{q}-k}
M^{pp,\sing}_{kk'{q}}G_{k'}G_{q-k'}f^\sing_{\nu'\nu{\omega}},\\
{\Sigma}^{\firr,\trip}_k=-&\frac{3}{4}\sum_{k'{q}}{G}_{{q}-k}
M^{pp,\trip}_{kk'{q}}G_{k'}G_{q-k'}f^\trip_{\nu'\nu{\omega}}.
\end{align}
\end{subequations}
The right diagram in Fig.~\ref{fig:twoboson} shows a two-boson exchange
taken into account by the vertex $M^{ph}$~\footnote{
Inside the Schwinger-Dyson Eq.~\eqref{eq:sigma} a vertical ladder diagram,
as depicted on the right of Fig.~\ref{fig:twoboson},
can be mapped to a horizontal one via the crossing-symmetry.}.

In the spirit of the original fluctuation diagnostic~\cite{Gunnarsson15},
it is possible to assign a bosonic argument to each self-energy component, that is $\Sigma_{kq}$,
which allows to find the wave vector and energy of the dominant bosonic contributions to the self-energy.
To do this, we can simply omit the summation over the bosonic label $q=(\qv,\omega)$.

\section{Numerical results}\label{sec:results}
We present results for the self-energy from the boson exchange parquet solver (BEPS).
The lattice size is fixed to $32\times32$ sites. For the truncated unity approximation~\cite{Eckhardt20}
of the residual vertex $\Phi^\firr$ \textit{we use only one form factor}.
Thanks to the fast convergence of BEPS with the number of form factors~\cite{Krien20},
the higher form factors are irrelevant at half-filling and $U/t=2$.

\subsection{Effect of self-consistency}
\begin{figure}
    \begin{center}
        \includegraphics[width=0.5\textwidth]{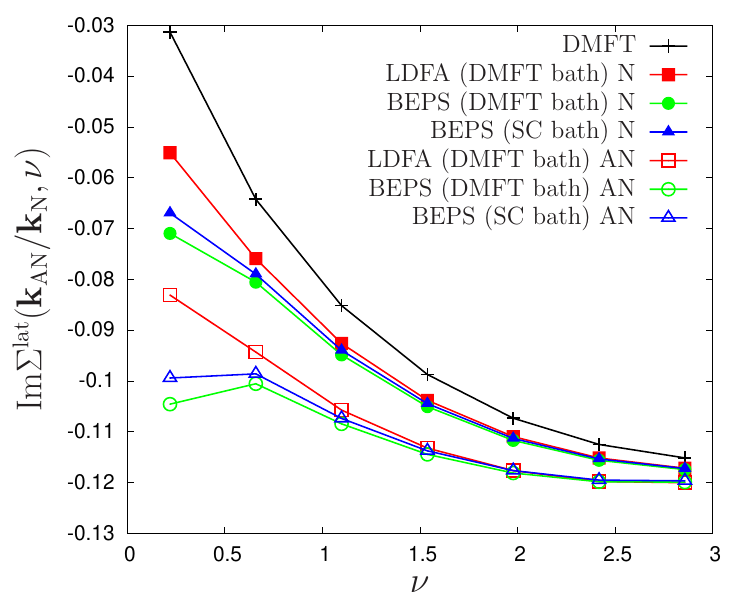}
    \end{center}
    \vspace{-0.5cm}
    \caption{\label{fig:sc} Self-energy of different methods at node (full symbols) and antinode (open symbols) for $T=0.07$.
    }
\end{figure}

First, we evaluate the BEPS self-energy at $T/t=0.07$
and examine the effect of the bath self-consistency.
Figure~\ref{fig:sc} shows that using the DMFT hybridization [i.e., applying the prescription~\eqref{eq:sc_dmft}]
the antinode already shows a clear non-Fermi liquid signature at this temperature.
This implies that the non-self-consistent BEPS has a slightly
stronger tendency to the pseudogap formation than the
numerically exact diagrammatic Monte Carlo (DiagMC),
where the antinode turns insulating near $T/t=0.065$~\cite{Schaefer20}.
For comparison we also show the momentum-independent DMFT self-energy
and the self-energy from the ladder dual fermion approach (LDFA,~\cite{Hafermann09}),
which does not show the nodal/antinodal differentiation at this temperature.

\begin{figure}
    \begin{center}
        \includegraphics[width=0.5\textwidth]{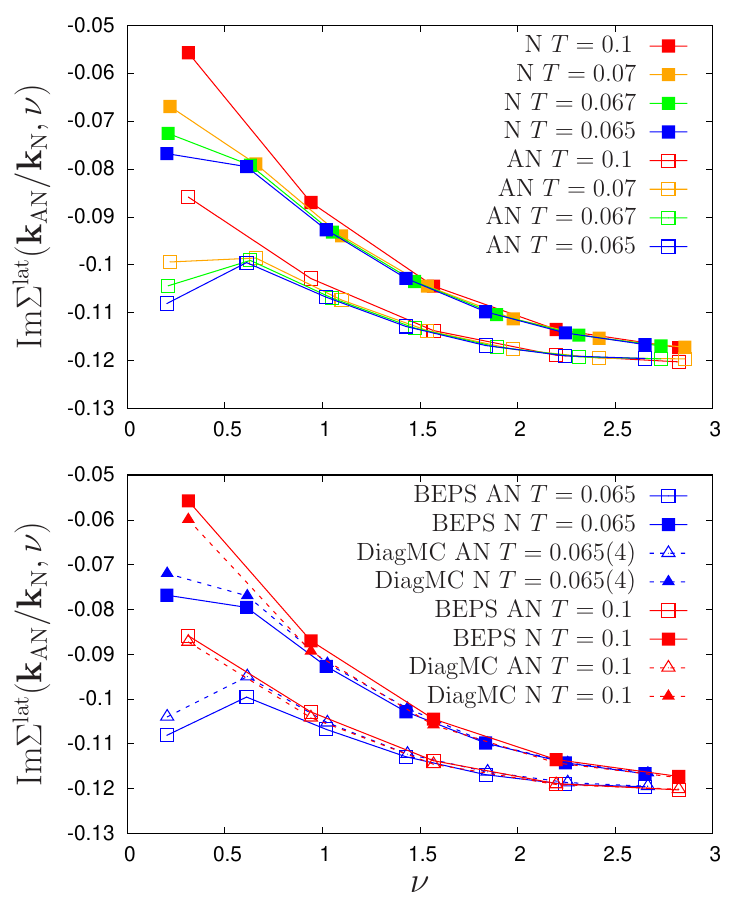}
    \end{center}
    \vspace{-0.5cm}
    \caption{\label{fig:diagmc} Top: BEPS self-energy
    at node (full symbols) and antinode (open symbols) for different temperatures.
    Bottom: Comparison with the DiagMC result of Ref.~\cite{Schaefer20}.
    }
\end{figure}

It is plausible that a self-consistent adjustment of the local impurity model can improve the result.
Indeed, the self-consistency condition~\eqref{eq:sc_df} for the dual Green's function, $G_\text{loc}=0$,
relaxes the tendency of the BEPS self-energy to the pseudogap formation
and turns the self-energy at the lowest Matsubara frequencies
slightly back toward the metallic direction (see Fig.~\ref{fig:sc}).
Apparently, the bath self-consistency has a sizable effect in the pseudogap regime
and should be applied in a quantitative comparison with an exact benchmark.
Henceforth, we use the prescription~\eqref{eq:sc_df} (outer self-consistency).
Practically, we converge at low temperature via annealing,
using the self-consistent hybridization determined at a slightly higher temperature.
Note that the self-consistency implies that the Hartree-Fock contribution to the self-energy vanishes,
\begin{align}
\Sigma^\text{HF}_\nu=\sum_{\nu'}G_{\text{loc}}(\nu')f^\ch_{\nu'\nu,\omega=0}=0.\label{eq:hf_sc}
\end{align}
We apply the same prescription~\eqref{eq:sc_df} also for our LDFA calculations~\footnote{
Due to the different feedback of BEPS and LDFA on the impurity model~\eqref{eq:aim} their self-consistent bath $h_\nu$ is not the same. Strictly speaking, this implies that the {\sl dual} quantities of these approximations
can not be compared quantitatively (concerning Figs.~\ref{fig:sigma_kq} and~\ref{fig:hedin_path}), however, this does not affect our qualitative discussion.
}.

\subsection{Quantitative comparison with DiagMC}
For a quantitative comparison with the numerically exact diagrammatic Monte Carlo results of Ref.~\cite{Schaefer20}
we calculate the self-energy using BEPS for $0.1\geq T/t\geq0.065$, see upper panel of Fig.~\ref{fig:diagmc}.
The sequence shows that the nodal/antinodal dichotomy develops in this temperature range.
In the lower panel we compare for $T/t=0.1$ and $T/t=0.065(4)$ to DiagMC.
The BEPS self-energy is in good quantitative agreement with DiagMC and shows
a consistent nodal/antinodal differentiation, overall in better agreement with DiagMC than a variety of
approximate methods benchmarked in Ref.~\cite{Schaefer20}.

\subsection{Self-energy decomposition}\label{sec:fdiag1}
We decompose the BEPS self-energy according to Sec.~\ref{sec:fdiag}.
We begin with the SBE decomposition which gives rise to the components of the dual self-energy in Eq.~\eqref{eq:fdiag_sbe}.
Inserting the decomposition into the expression~\eqref{eq:sigma_lat} for the lattice self-energy we arrive at,
\begin{align}
\Sigma^\text{lat}_k
=&\Sigma^\text{imp}_\nu+\frac{\Sigma^{\firr}_k+\Sigma^\ch_k+\Sigma^\sz_k+\Sigma^\sing_k+{\Sigma}^{\text{bare}}_k}{1+g_\nu\Sigma_k}\notag\\
=&\Sigma^\text{imp}_\nu+\Sigma'^{\firr}_k+\Sigma'^\ch_k+\Sigma'^\sz_k+\Sigma'^\sing_k+{\Sigma}'^{\text{bare}}_k
\label{eq:sigma_lat_sbe}
\end{align}
where we used that the Hartree-Fock self-energy vanishes [Eq.~\eqref{eq:hf_sc}].
The lattice self-energy $\Sigma^\text{lat}$ is thus split into the local impurity self-energy $\Sigma^\text{imp}$
and five nonlocal components, each one divided by the same denominator $1+g_\nu\Sigma_k$,
which we absorbed in the second line of Eq.~\eqref{eq:sigma_lat_sbe} into definition of $\Sigma'$,
e.g., $\Sigma'^\ch_k=\Sigma^\ch_k/(1+g_\nu\Sigma_k)$.

\begin{figure}
    \begin{center}
        \includegraphics[width=0.5\textwidth]{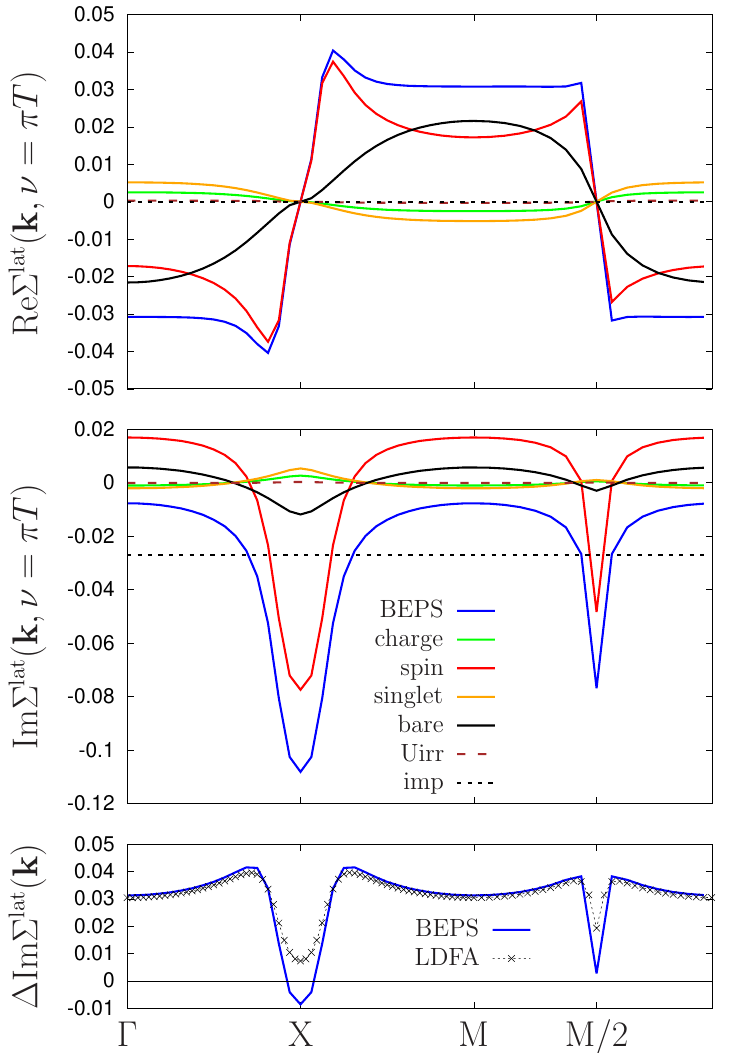}
    \end{center}
    \vspace{-0.5cm}
    \caption{\label{fig:path_sbe_t0.065} Fluctuation diagnostic of the real (top) and imaginary (center)
    part of the BEPS \textit{lattice} self-energy at the first Matsubara frequency,
    along the high-symmetry path ($T=0.065$).
    The solid blue line shows the full self-energy, a dashed line the local self-energy of
    the self-consistent AIM. The other lines indicate the various components $\Sigma'$ defined in Eq.~\eqref{eq:sigma_lat_sbe}.
    The bottom panel shows the non-Fermi-liquid marker 
    $\Delta\text{Im}\Sigma^\text{lat}$ (see text),
    black symbols indicate the LDFA.
    }
\end{figure}

In Fig.~\ref{fig:path_sbe_t0.065}, the various components are drawn at the first Matsubara frequency
along a high-symmetry path in momentum space for $T=0.065$.
The leading nonlocal contributions to the self-energy are the single-(spin)boson exchange,
$\Sigma'^\sz$, and the bare contribution ${\Sigma}'^{\text{bare}}$, which have, in general, the same sign.
The next largest contributions are due to single-boson exchange in the charge and singlet channels.
Consistent with the observation in Ref.~\cite{Gunnarsson16} they have the opposite sign of $\Sigma'^\sz$. At half filling, theses contributions are very small due to the suppression of charge and particle-particle fluctuations in this parameter regime.

The contribution $\Sigma'^{\firr}$ due to multi-boson exchange is negligible
in the weak coupling regime considered here.
We show the decomposition of $\Sigma'^{\firr}$ according to Eq.~\eqref{eq:fdiag_firr}
in Appendix~\ref{app:mbe}.

The bottom panel of Fig.~\ref{fig:path_sbe_t0.065} shows 
the difference between the self-energy at the first and the second Matsubara frequency,
$\Delta\text{Im}\Sigma^\text{lat}(\kv)=
\text{Im}\Sigma^\text{lat}(\kv,\pi T)-\text{Im}\Sigma^\text{lat}(\kv,3\pi T)$.
For a metal, $\Delta\text{Im}\Sigma^\text{lat}>0$ for momenta $\mathbf{k}$ at the Fermi level.
The crossover to the (insulating) non-Fermi-liquid
regime is then roughly indicated by $\Delta\text{Im}\Sigma^\text{lat}$
crossing 0~\cite{Schaefer15,Schaefer20}.
At the chosen temperature $T=0.065$, we find that the BEPS
self-energy features already a pseudogap behavior, i.e.,  a non-Fermi-liquid-type self-energy at the antinode ($\Delta\text{Im}\Sigma^\text{lat}<0$ at the $X$ point) while the node ($M/2$) is still metallic. On the contrary, the corresponding results for the LDFA (black symbols, cf. Appendix~\ref{app:ladder}) indicate that we are still in the metallic regime and the pseudogap opens at a lower temperature than in BEPS or DiagMC~\cite{Schaefer20}.

Nevertheless, in both dual fermion approximations the non-Fermi-liquid behavior arises from the
single-spinboson exchange, $\Sigma'^\sz$.
We therefore discuss in the following only the diagram on the left-hand side
of Fig.~\ref{fig:twoboson} with the boson flavor $\alpha=\sz$, that is,
we can safely ignore the contributions of (single) charge and singlet bosons
as well as any sort of multi-boson exchange such as the right diagram in Fig.~\ref{fig:twoboson}.

\begin{figure}
    \begin{center}
    \begin{tikzpicture}
      \node[anchor=south west,inner sep=0] (image1) at (0,0)    {\includegraphics[width=0.5\textwidth]{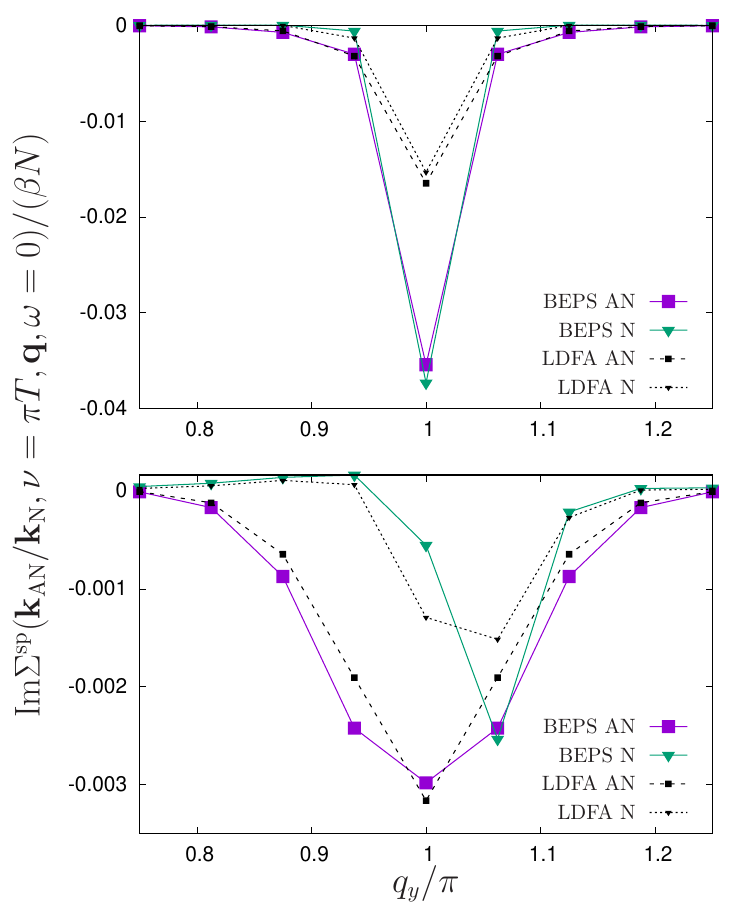}};
      \draw (3,6.5) node{\large$q_x=\pi$};
      \draw (3,1.5) node{\large$q_x=\frac{15}{16}\pi$};
    \end{tikzpicture}
    \end{center}
    \vspace{-0.5cm}
    \caption{\label{fig:sigma_kq}
    $\qv$-resolved fluctuation diagnostic of $\Sigma^\sz$ at $T=0.065$.
    Solid and dashed lines show BEPS ($N=32^2$) and LDFA ($N=64^2$), respectively.
    For comparison the LDFA result is mapped to the smaller lattice.
    The antinode is affected more strongly by incommensurate fluctuations than the node.
    }
\end{figure}

\subsection{Fluctuation diagnostic}
We perform a fluctuation diagnostic~\cite{Gunnarsson15} of the single-spinboson exchange,
which reveals the origin of the nodal/antinodal dichotomy in the pseudogap regime.
To this end, we omit the summation over $q=(\qv,\omega)$ in Eq.~\eqref{eq:sbe_sp}, that is,
\begin{align}
{\Sigma}^{\sz}_{kq}=-&\frac{3}{2}\sum_{k'}{G}_{k+q}({\Delta}^{ph,\sz}_{kk'q}\!-\!U^\sz)
G_{k'}G_{k'+q}f^\sz_{\nu'\nu\omega}.\notag
\end{align}
We draw the quantity
\begin{align}
\frac{1}{\beta N}\text{Im}{\Sigma}^{\sz}(\kv\!=\!\kv_N/\kv_{AN},\nu=\!\pi T\!,\qv\!=\!(q_x,q_y),\omega\!=\!0)\notag
\end{align}
in Fig.~\ref{fig:sigma_kq} as a function of $q_y$ for fixed $q_x=\pi$ (top panel)
and $q_x=\frac{15}{16}\pi$ (bottom panel).
This allows us to analyze how much spin fluctuations with a certain wave vector $(q_x,q_y)$ contribute to
the nodal/antinodal self-energy at the first Matsubara frequency~\footnote{
For a corresponding LDFA result for the $t,t'$-Hubbard model compare Fig. 5 of Ref.~\cite{Arzhang20}.}.
Remarkably, our BEPS results show that spin fluctuations with the exact commensurate nesting vector $\qv=(\pi,\pi)$
contribute equally (or even slightly stronger) to the node than to the antinode
(see top panel of Fig.~\ref{fig:sigma_kq} at $q_y/\pi=1$).
The larger self-energy at the antinode then originates from incommensurate momenta \textit{near} $(\pi,\pi)$ which are weaker for the node as can be seen in the lower panel of Fig.~\ref{fig:sigma_kq}.

Remarkably, there is another interesting feature for $q_x\!=\!\frac{15}{16}\pi\!=\!\pi\!-\!\frac{1}{16}\pi$: While for the antinode the dominant contribution to the self-energy remains at $q_y\!=\!\pi$, for the node we observe the maximum at $q_y\!=\!\frac{17}{16}\pi\!=\!\pi\!+\!\frac{1}{16}\pi$. Interestingly, the same behavior is found already for the LDFA self-energy {albeit with overall smaller values since the LDFA predicts a more metallic behavior at this temperature (and, hence, a smaller self-energy).}

\begin{figure}
    \begin{center}
    \begin{tikzpicture}[domain=0:2, scale = 0.7]
    \draw[black, line width = 0.50mm]   plot[smooth,domain=-1:1] (\x, {-\x});
    \draw[black, line width = 0.50mm]   plot[smooth,domain=2:4] (\x, {(\x-3)});
    \draw[black, line width = 0.30mm, <->] (0,.2) -- (3,.2);
    \node[anchor=east] at (2,.7) {$\qv$};
    \node[anchor=east] at (0,-.8) {(a)};
    \begin{scope}[shift={(5.5,0)}]
    \draw[black, line width = 0.50mm]   plot[smooth,domain=-1:0] (\x, {+(\x)*(\x)});
    \draw[black, line width = 0.50mm]   plot[smooth,domain=0:1]  (\x, {-\x*\x});
    \draw[black, line width = 0.50mm]   plot[smooth,domain=2:3] (\x, {-(\x-3)*(\x-3)});
    \draw[black, line width = 0.50mm]   plot[smooth,domain=3:4]  (\x, {+(\x-3)*(\x-3)});
    \draw[black, line width = 0.30mm, <->] (0,.2) -- (3,.2);
    \node[anchor=east] at (2,.7) {$\qv$};
    \node[anchor=east] at (0,-.8) {(b)};
    \end{scope}
    \end{tikzpicture}
    \end{center}
    \vspace{-0.5cm}
    \caption{\label{fig:sketch}
    Sketch of the nodal/antinodal dichotomy.
    The nesting vector $\qv=(\pi,\pi)$ connects equivalent regions
    of the square lattice dispersion $\varepsilon_\kv\!\propto\!\cos(k_x)\!+\!\cos(k_y)$ on the Fermi surface.
    (a) Finite first derivative (e.g., nodes): Scatterings require a fine-tuning of $|\qv|$.
    (b) Vanishing first derivative (antinodes): Scatterings are insensitive to small changes in $|\qv|$.
    }
\end{figure}
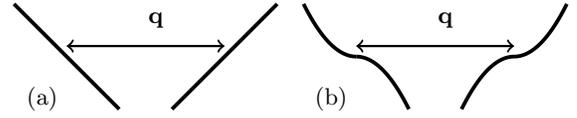

The similarity between the parquet and the ladder results suggests that the effects discussed above might be understood by means of an even simpler approach. Indeed, as it is discussed in detail in Appendix~\ref{app:spinfermion}, a spin-fermion-like calculation of the self-energy shows that the different line shapes for antinode and node in Fig.~\ref{fig:sigma_kq} originate in the van Hove singularity of the square lattice. The actual mechanism at work is illustrated in Fig.~\ref{fig:sketch}: The scattering between the flat regions around the saddle points~\cite{Irkhin2001,Irkhin2002} of the dispersion relation at the Fermi surface (corresponding to the antinode) does not require a fine-tuning of the bosonic momentum to the nesting vector $\mathbf{q}\!=\!(\pi,\pi)$~\cite{Metzner12} [see Fig.~\ref{fig:sketch}(b)]. On the contrary, for momenta  with a finite Fermi velocity [Fig.~\ref{fig:sketch}(a)] a significantly enhanced scattering rate can be expected only for the {\em exact} nesting vector. For the corresponding explicit analytical justification of this argument we refer the reader to Appendix~\ref{app:spinfermion}.

\begin{figure}
    \begin{center}
        \begin{tikzpicture}
        \node[anchor=south west,inner sep=0] (image1) at (0,0)    {\includegraphics[width=0.5\textwidth]{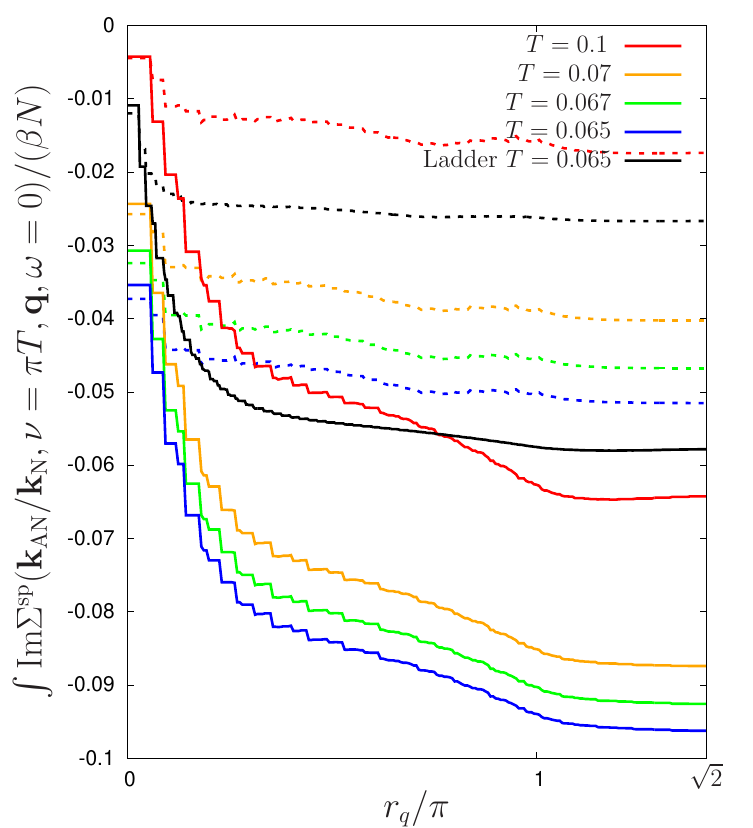}};
        \draw (6,6.8) node{\large N};
        \draw (6,3.5) node{\large AN};
        \end{tikzpicture}
    \end{center}
    \vspace{-0.5cm}
    \caption{\label{fig:sigma_kq_int} Self-energy component $\Sigma^\sz_{kq}$,
    integrated over a circle with radius $r_q$ centered at $\qv=(\pi,\pi)$
    for different temperatures at $\omega\!=\!0$.
    At the node (dashed lines) the integral is largely determined by the value at $r_q=0$,
    at the antinode (solid lines) incommensurate spin fluctuations
    contribute significantly. Black lines show LDFA for $T=0.065$.
    }
\end{figure}

To get more insights into the origin of the nodal/antinodal dichotomy of the self-energy,
we integrate (sum) $\frac{1}{\beta N}{\Sigma}^{\sz}_{kq}$ with respect to $\qv$
over the area covered by a circle with radius $r_q$, centered at $(\pi,\pi)$.
The result is shown in Fig.~\ref{fig:sigma_kq_int}
as a function of $r_q$ for $T=0.065, 0.067, 0.07, 0.1$.
Since ${\Sigma}^{\sz}_{kq}$ is known only at discrete momenta,
the integral grows step-wise as $r_q$ is increased.
Indeed, at the nodal point the integral is determined almost entirely by spin fluctuations with momentum $(\pi,\pi)$.
In contrast, at the antinode incommensurate momenta $(\pi\pm\varepsilon_x,\pi\pm\varepsilon_y)$
with sizable $\varepsilon_{x/y}\lesssim0.2\pi$ contribute significantly to the integral.
In fact, the incommensurate spin fluctuations are responsible for the nodal/antinodal dichotomy. Also this aspect is explained by the spin-fermion-like calculation in Appendix~\ref{app:spinfermion}.

\begin{figure}
    \begin{center}
        \begin{tikzpicture}
        \node[anchor=south west,inner sep=0] (image1) at (0,0)    {\includegraphics[width=0.5\textwidth]{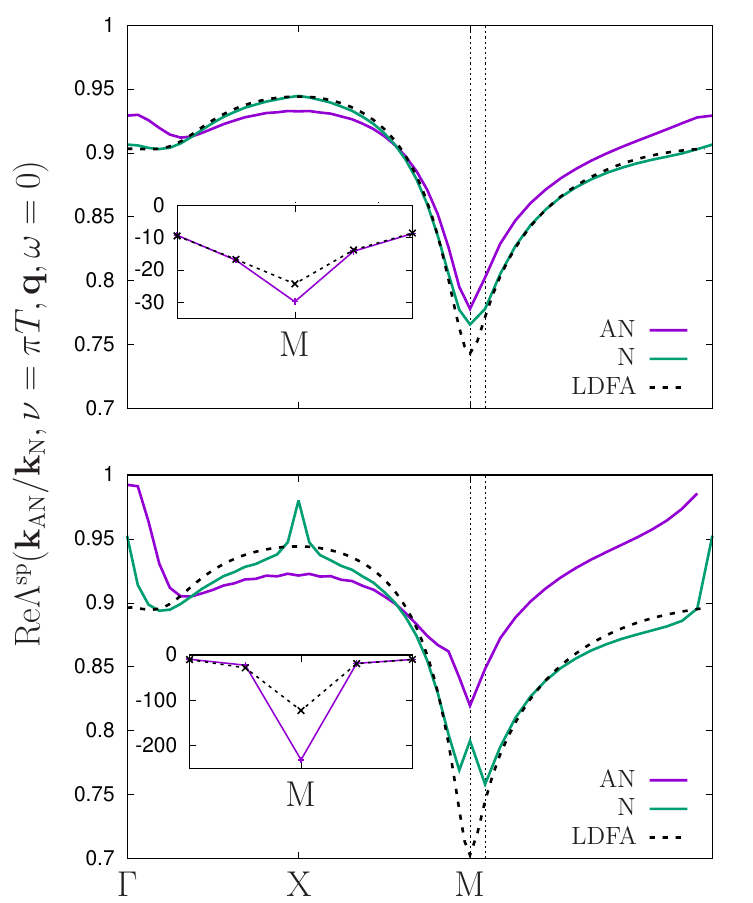}};
        \node[anchor=east] at (7.5,10) {$T=0.1$};
        \node[anchor=east] at (7.5,4.7) {$T=0.065$};
        \end{tikzpicture}
    \end{center}
    \vspace{-0.5cm}
    \caption{\label{fig:hedin_path} Main panels:
    Fermion-spinboson coupling as a function of
    $\qv$ along the high-symmetry path, above (top) and below (bottom)
    the temperature where the antinodal BEPS self-energy turns insulating.
    Dashed curves show the ladder approximation.
    Vertical lines indicate (in)commensurate momenta in Fig.~\ref{fig:hedin_azimuth}.
    Insets: Dual screened interaction $W^\sz(\qv,\omega\!=\!0)$
    of BEPS (solid) and LDFA (dashed) near $(\pi,\pi)$.
    }
\end{figure}


\begin{figure}
    \begin{center}
    \begin{tikzpicture}
    \node[anchor=south west,inner sep=0] (image1) at (0,0)
     {\includegraphics[width=0.5\textwidth]{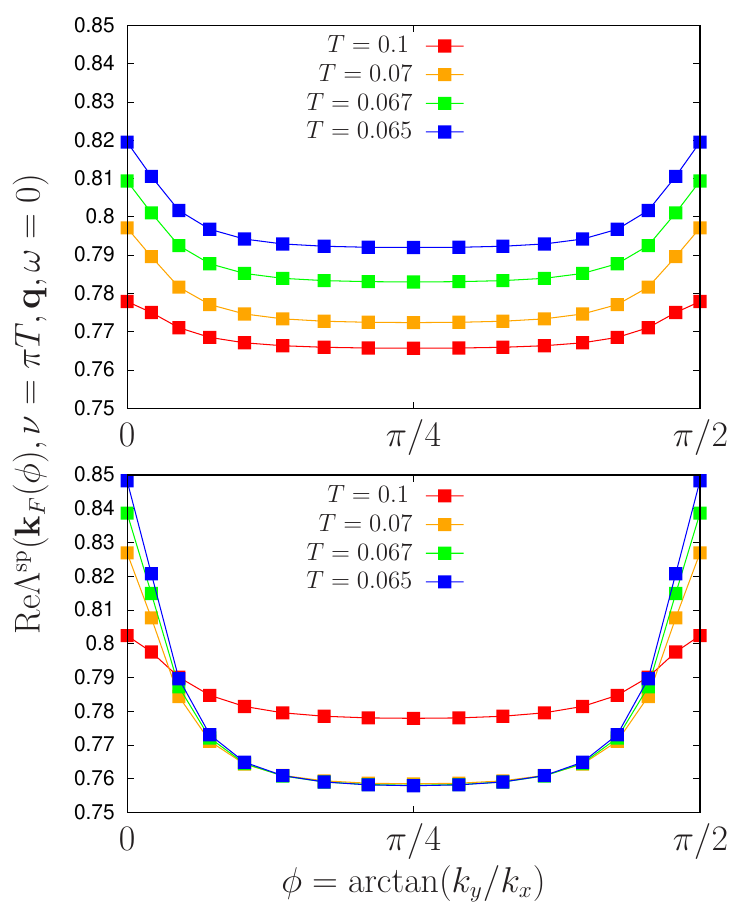}};
        \node[anchor=east] at (5.5,8.5) {$\qv=(\pi,\pi)$};
        \node[anchor=east] at (5.5,3) {$\qv=\frac{15}{16}(\pi,\pi)$};
        \end{tikzpicture}
    \end{center}
    \vspace{-0.5cm}
    \caption{\label{fig:hedin_azimuth} Coupling of fermions with Fermi vector $\kv_F$, pointed in direction of
    the angle $\phi$, to commensurate (top) and incommensurate (bottom) spin fluctuations.
    A strong nodal/antinodal dichotomy, related to the pseudogap formation,
    develops only with respect to {\sl in}commensurate spin fluctuations.
    The antinode ($\phi=0,{\pi}/{2}$) is coupled more strongly
    to these fluctuations than the node ($\phi={\pi}/{4}$).
    }
\end{figure}

\subsection{Fermion-boson coupling}
To further investigate how the node and antinode respond differently to spin fluctuations
we examine the fermion-spinboson coupling (Hedin vertex),
\begin{align}
\Lambda^\sz(\kv\!=\!\kv_N/\kv_{AN},\nu\!=\!\pi T,\qv,\omega\!=\!0).
\end{align}
In Fig.~\ref{fig:hedin_path} we show this quantity as a function of $\qv$ for $T=0.1$ and $T=0.065$, outside and inside the pseudogap regime, respectively.
In the latter case, this vertex also exhibits a nodal/antinodal dichotomy where,
interestingly, in the nodal direction it develops spikes directed upwards at commensurate bosonic momenta, $\qv=(0,0), (\pi,0),$ and $(\pi,\pi)$,
{which connect the node with itself and other nodal points, respectively}.
Due to the spike at $(\pi,\pi)$, the coupling of the node and antinode to spin fluctuations with this wave vector is comparable. This is shown, in more detail, in the upper panel of Fig.~\ref{fig:hedin_azimuth}, where $\Lambda^\sz(\kv_F(\phi),\nu=\pi T,\qv=(\pi,\pi),\omega=0)$
is drawn for points $\kv_F(\phi)$ on the Fermi surface in the direction of $\phi=\arctan(k_y/k_x)$.
The coupling to the $(\pi,\pi)$-spin fluctuations (upper panel) depends only weakly on $\phi$ and looks qualitatively similar at high and low temperature.
However, changing the bosonic momentum by only one unit along the Brillouin zone diagonal to $\frac{15}{16}(\pi,\pi)$ the result differs qualitatively (lower panel):
Upon opening of the pseudogap ($T\lesssim0.07$), the coupling of antinodal fermions ($\phi=0,\pi/2$)
to incommensurate spin fluctuations is enhanced compared to the node ($\phi=\pi/4$).
This dichotomy concerns only low-energetic fermions with $|\nu|\lesssim\pi T$.

Note that such a difference between the node and the antinode can not be found in the LDFA calculations (dashed lines in Fig.~\ref{fig:hedin_path}) as the fermion-spinboson coupling does not depend on the momentum $\mathbf{k}$ (see, for example, Ref.~\cite{Krien18}). {One can also see, that the coupling to antiferromagnetic fluctuations ($M$ point) is weaker in the LDFA with respect to the BEPS results. Together with the reduced strength of the spin fluctuations in LDFA
[see $W^{\text{sp}}(\mathbf{q},\omega\!=\!0)$ in the insets of Fig.~\ref{fig:hedin_path}],
this results in an overall smaller self-energy as has been discussed in the previous section.}

Let us turn our attention to the interesting spike at the $X$-point $\mathbf{q}\!=\!(\pi,0)$ for the fermion-spinboson coupling vertex at the nodal momentum in the lower panel of Fig.~\ref{fig:hedin_path}. In principle, this peak would indicate an enhanced coupling of spin stripe fluctuations to the nodal fermions. However, such fluctuations are suppressed at half-filling which suggests that this stronger value of the coupling is irrelevant for the self-energy. However, the actual origin and the physical meaning of this feature requires further investigations.

\begin{figure}
    \begin{center}
    \begin{tikzpicture}
    \node[anchor=south west,inner sep=0] (image1) at (0,0)
        {\includegraphics[width=0.5\textwidth]{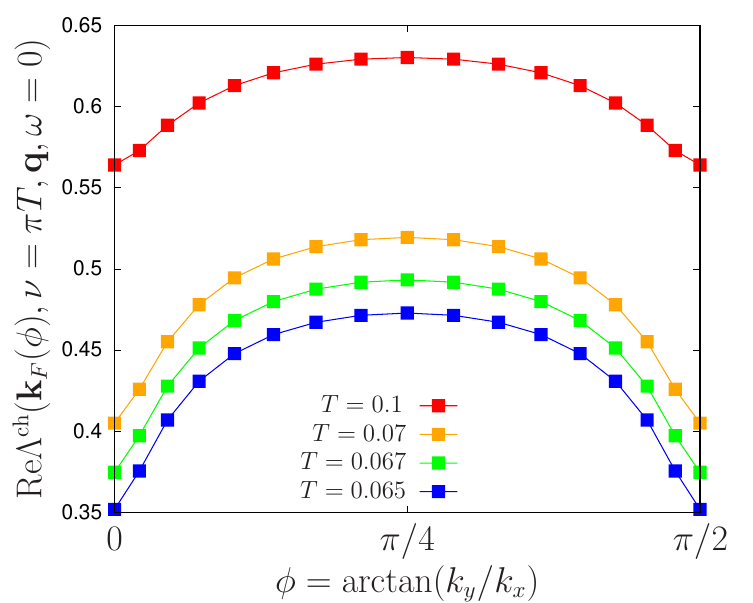}};
        \node[anchor=east] at (5.5,5) {$\qv=(\pi,\pi)$};
         \end{tikzpicture}
    \end{center}
    \vspace{-0.5cm}
    \caption{\label{fig:hedin_azimuth_ch} 
    Coupling of fermions with Fermi vector $\kv_F$ pointed in direction of
    the angle $\phi$ to charge fluctuations with commensurate momentum $\qv=(\pi,\pi)$.
    In the pseudogap regime the fermions are strongly screened from charge fluctuations.
    }
\end{figure}

Finally, we briefly comment on the role of charge degrees of freedom. Figure~\ref{fig:hedin_azimuth_ch} shows the coupling $\Lambda^\ch$ of fermions to such charge fluctuations. In the pseudogap regime, this vertex is considerably smaller than the corresponding coupling to spin fluctuations. 
In combination with the already strongly suppressed charge susceptibility, this further reduces the impact of charge fluctuations on the self-energy (see also Ref.~\cite{Schmalian99}).
Furthermore, at particle-hole symmetry singlet fluctuations contribute to the self-energy
exactly two times as much as the charge fluctuations
(see yellow and green lines in Fig.~\ref{fig:path_sbe_t0.065}).
Therefore, in LDFA it would be more consistent to neglect the charge and singlet fluctuations on equal footing.

\section{Conclusions}\label{sec:conclusions}
We presented a fluctuation diagnostic of the nodal/antinodal dichotomy in the half-filled Hubbard model
on the simple square lattice at weak coupling.
To this end, we employed a novel method, the boson exchange parquet solver (BEPS) for dual fermions,
presented in the accompanying Ref.~\cite{Krien20}.
The self-energy calculated in this method is in good quantitative
agreement with the numerically exact
diagrammatic Monte Carlo (DiagMC) results of Ref.~\cite{Schaefer20}
and shows a consistent nodal/antinodal
differentiation in the pseudogap regime.

We demonstrated that the vertex decomposition used in the BEPS method implies
a fluctuation diagnostic~\cite{Schaefer2020a} of the self-energy that does not rely on the Fierz ambiguity,
which is the basis for the original fluctuation diagnostic introduced in Ref.~\cite{Gunnarsson15}.
The key for removing the Fierz ambiguity is to represent vertex corrections in terms
of the screened interaction, which has a unique flavor~\cite{Krien19-2,Denz19}.
Only the bare interaction diagram is intrinsically ambiguous
and can not be assigned to a unique channel.
For weak coupling the bare interaction diagram is important and 
its contribution to the self-energy is smooth as a function of momentum.
At low temperature the feedback of the spin fluctuations adds
sharp features to the self-energy.

We found that the nodal/antinodal dichotomy does not originate
in spin fluctuations with the commensurate wave vector $(\pi,\pi)$.
Rather, {\sl in}commensurate spin fluctuations contribute significantly
to the antinodal self-energy but not to the nodal one.
In fact, the dichotomy is entirely the result of slightly incommensurate fluctuations,
which is confirmed by a spin-fermion-like calculation (cf. Appendix~\ref{app:spinfermion}).
It is intriguing to consider the role of incommensurate fluctuations also in the doped $t$-$t'$-Hubbard model, see, e.g., Ref.~\cite{Wu17}.
A fluctuation diagnostic based on the dynamical cluster approximation (DCA,~\cite{Gunnarsson15})
identified the $(\pi,\pi)$-fluctuations as the origin of the momentum differentiation, however,
the incommensurability relevant at least in our study is smaller than the patch size of $8$-site DCA.

Finally, we showed that in our BEPS calculations the nodal/antinodal dichotomy also becomes manifest in the fermion-spinboson coupling, which develops a rich dependence on the fermionic momentum $\mathbf{k}$ in the pseudogap regime.
Interestingly, nodal and antinodal fermions couple similarly to
commensurate spin fluctuations with wave vector $(\pi,\pi)$ at high and low temperature,
whereas the coupling to slightly incommensurate spin fluctuations exhibits the dichotomy. Such features {\sl can not} be captured by any kind of ladder approximation since the corresponding fermion-spinboson vertex becomes $\mathbf{k}$-independent in this case.

Our calculations were simplified by the particle-hole symmetry of the half-filled Hubbard model. However, an exact particle-hole symmetry is only rarely found in realistic materials. Hence, in spite of some technical difficulties such as the difference in the lattice and impurity densities, we are already working to extend our methods to the doped case in order to study, e.g., the physics of the high-temperature superconducting cuprates.

\acknowledgments
We thank F. \v{S}imkovic for providing the DiagMC data and K. Held, A. Kauch, T. Sch\"afer, and A. Toschi for useful discussions. We acknowledge support from the 
Slovenian Research Agency through Project No. N1-0088 (F.K.), from the Austrian Science Fund (FWF) through Projects No. P32044 and No. P30997 (F.K.), from the Deutsche Forschungsgemeinschaft (DFG) through Project No. 407372336 (G.R.), and from the European Research Council through the Synergy Grant No. 854843 - FASTCORR (A.I.L.).

\appendix
\section{Multi-boson exchange}\label{app:mbe}

\begin{figure}
    \begin{center}
        \includegraphics[width=0.5\textwidth]{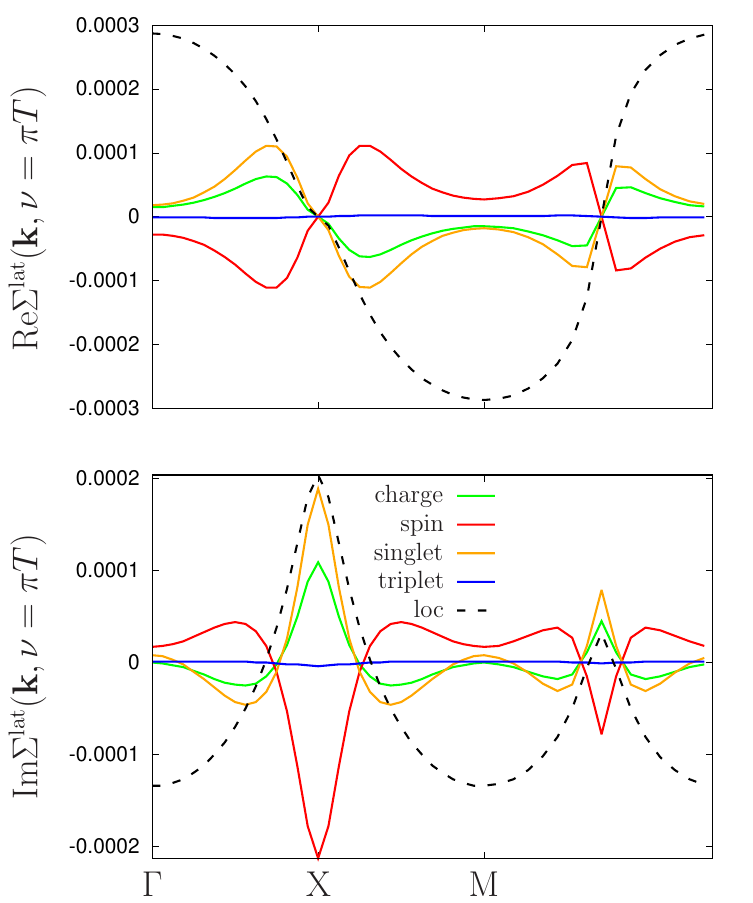}
    \end{center}
    \vspace{-0.5cm}
    \caption{\label{fig:path_mbe_t0.065}Self-energy components due to multi-boson exchange,
    which are negligible compared to the single-boson exchange shown in Fig.~\ref{fig:path_sbe_t0.065}.
    The dashed line indicates the contribution of the local residual vertex $\varphi^\firr$,
    other lines indicate the four channels of the parquet equations.
    }
\end{figure}

We show in Fig.~\ref{fig:path_mbe_t0.065} the decomposition of $\Sigma'^{\firr}_k$ via Eq.~\eqref{eq:fdiag_firr}
into charge, spin, singlet, and triplet components,
\begin{align}
\Sigma'^{\firr}_k=&\Sigma'^{\firr,\text{loc}}_k+\Sigma'^{\firr,\ch}_k+\Sigma'^{\firr,\sz}_k\notag\\
+&\Sigma'^{\firr,\sing}_k+\Sigma'^{\firr,\trip}_k.
\label{eq:sigma_lat_mbe}
\end{align}
The contribution $\Sigma'^{\firr,\text{loc}}$ is due to the local vertex $\varphi^\firr$,
the first term on the right-hand-sides of Eqs.~\eqref{eq:parquet_ch} and~\eqref{eq:parquet_sp}.
Again, the prime indicates that each component of Eq.~\eqref{eq:fdiag_firr} was divided by $1+g_\nu\Sigma_k$.
As it was observed previously in related parquet decomposition calculations\cite{Gunnarsson16,Rohringer2013a} and also for the single-boson exchange contributions (see Sec.\ref{sec:fdiag1}), the particle-particle and charge fluctuations screen the (otherwise overestimated) spin fluctuations. However, for the small value of $U\!=\!2$ considered here, all these terms are about three orders of magnitude smaller than the corresponding single-boson exchange contributions.

\section{Ladder approximation}\label{app:ladder}
In our comparison with the ladder dual fermion approach (LDFA,~\cite{Hafermann09}) we
calculate the dual self-energy $\Sigma$ consistent with Ref.~\cite{Otsuki14}.
We impose the self-consistency condition~\eqref{eq:sc_df}.
The lattice size is fixed to $64\times64$.

The LDFA self-energy is not usually expressed in terms of boson exchange.
However, we can obtain the Hedin vertex $\Lambda^\alpha(\nu,q)$
and the screened interaction $W^\alpha(q)$ for $\alpha=\ch,\sz$ corresponding to the LDFA
similar to Ref.~\cite{Krien19} where this is done for the DMFT approximation~\cite{Katanin09}.
The only difference is that the dual Green's functions are now dressed with the LDFA self-energy.
Then, from $\Lambda$ and $W$ we form the SBE vertex,
\begin{align}
\Delta^{ph,\alpha}_{\nu\nu'q}=\Lambda^\alpha(\nu,q)W^\alpha(q)\Lambda^\alpha(\nu',q),
\end{align}
which is thus independent of $\kv,\kv'$, and obtain the self-energy component
$\Sigma^\sz$ analogous to Eq.~\eqref{eq:sbe_sp}.

\section{Spin-fermion-like calculations}\label{app:spinfermion}

In this section, we discuss the dichotomy between the nodal and the antinodal momentum in the self-energy from a simplified spin-fermion- (or a paramagnon-)like perspective. In particular, we will show how the saddle point in the bare dispersion at the antinode, which is responsible for the van Hove singularity in the non-interacting density of states, leads to a momentum differentiation in the self-energy as found in Fig.~\ref{fig:sigma_kq}.
In this Appendix we denote the Green's function and spin susceptibility of the lattice as $G$
and $\chi^\sz$, respectively.

We start from a spin-fermion-like expression~\cite{Abanov2003,Katanin09} for the equation of motion
[$\boldsymbol{\pi}=(\pi,\pi)$],
\begin{equation}
    \label{equ:sigmaspinfermion}
    \Sigma(\mathbf{k},\nu)\cong \sum_\mathbf{q}\underset{\chi^\sz(\mathbf{q},\omega=0)}{\underbrace{\frac{1}{(\mathbf{q}-\boldsymbol{\pi})^2+\xi^{-2}}}}\underset{G(\mathbf{k}+\mathbf{q},\nu)}{\underbrace{\frac{1}{\zeta(\mathbf{k}+\mathbf{q},\nu)-\varepsilon_{\mathbf{k}+\mathbf{q}}}}},
\end{equation}
which describes the scattering of an electron given by the Green's function $G(\mathbf{k},\nu)$ with an antiferromagnetic spin fluctuation represented by $\chi^\sz(\mathbf{q},\omega=0)$ in the Ornstein-Zernike form. In principle, the right-hand side of Eq.~\eqref{equ:sigmaspinfermion} should be also summed over $\omega$ but for finite temperatures the classical contribution  $\omega\!=\!0$ dominates.
We defined $\zeta(\mathbf{k}+\mathbf{q},\nu)\!=\!i\nu+\mu-\Sigma(\mathbf{k}+\mathbf{q},\nu)$ where the form of $\Sigma(\mathbf{k}+\mathbf{q},\nu)$ is not important for our argument (and it could be also set to 0). Let us point out that Eq.~(\ref{equ:sigmaspinfermion}) can be viewed in terms of the $GW$ theory~\cite{Hedin65} (cf. also Refs.~\cite{Krien19-2,Krien19-3})
where the Hedin vertex is set to 1, $W^\sz\propto\chi^\sz$
(and using the approximations for $\chi^\sz$ and $G$ mentioned above).

Obviously, momenta near $\mathbf{q}\!\approx\!\boldsymbol{\pi}$ yield the dominant contribution to the summation in Eq.~(\ref{equ:sigmaspinfermion}). Hence, after a shift of this integration variable, $\mathbf{q}\!\rightarrow\!\mathbf{q}\!+\!\boldsymbol{\pi}$, we expand the dispersion relation around $\mathbf{q}\!=\!\mathbf{0}$ for both the nodal $\mathbf{k}_{\text{N}}\!=\!(\frac{\pi}{2},\frac{\pi}{2})$ and the antinodal $\mathbf{k}_{\text{AN}}\!=\!(\pi,0)$ points:
\begin{align}
\label{equ:expanddispersion}
    \mathbf{k}_{\text{N}}:
    \quad\varepsilon_{\mathbf{k}+\mathbf{q}+\boldsymbol{\pi}}=&-2t(q_x+q_y)+O(\mathbf{q}^2),\\
    \mathbf{k}_{\text{AN}}:
    \quad\varepsilon_{\mathbf{k}+\mathbf{q}+\boldsymbol{\pi}}=&\;\;-t(q_x^2-q_y^2)+O(\mathbf{q}^3).
\end{align}
Here, we can already notice a crucial difference between the nodal ($\mathbf{k}_{\text{N}}$) and the antionodal ($\mathbf{k}_{\text{AN}}$) momentum: While the first features a linear term the second one exhibits a saddle point, i.e., the series expansion starts with the quadratic contribution.
We analyze the contributions to the sum in Eq.~(\ref{equ:sigmaspinfermion}) for small $\mathbf{q}\!\approx\!0$. For the nodal point we obtain
\begin{equation}
    \label{equ:qsumnodal}
    \Sigma(\mathbf{k}_{\text{N}},\nu)
    \cong\sum_\mathbf{q}\frac{1}{\mathbf{q}^2+\xi^{-2}}\frac{1}{\zeta+2t(q_x+q_y)}.
\end{equation}
In the spirit of a fluctuation diagnostic~\cite{Gunnarsson15},
we consider contributions of individual momenta $\qv$ to the sum:

At half-filling $\zeta$ is purely imaginary, therefore, the largest contribution arises from
the {commensurate} momentum $q_x\!=\!q_y\!=\!0$ (corresponding to $q_x\!=\!q_y\!=\!\pi$ before the shift of this variable). However, for a slightly {\sl in}commensurate momentum, e.g., $q_x\!=\!-\frac{\pi}{16}$, the second factor under the sum has its maximum at $q_y\!=\!+\frac{\pi}{16}$. In fact, this reduces the contribution of the first factor compared to $q_y\!=\!0$. In the first factor $\mathbf{q}$ enters quadratically, which is much smaller than the linear contribution $q_x+q_y$ in the second factor for $\mathbf{q}^2\ll1$. Hence, for $q_x\!=\!-\frac{\pi}{16}$ we expect the largest contribution from $q_y\!=\!+\frac{\pi}{16}$ as it can be indeed seen in the lower panel of Fig.~\ref{fig:sigma_kq}.

As a result, the contribution of incommensurate momenta to the nodal self-energy
is suppressed compared to the commensurate one, $\qv=0$.

On the contrary, for the antinodal point the expression for the self-energy becomes
\begin{equation}
    \label{equ:qsumantinodal}
    \Sigma(\mathbf{k}_{\text{AN}},\nu)
    \cong\sum_\mathbf{q}\frac{1}{\mathbf{q}^2+\xi^{-2}}\frac{1}{\zeta+t(q_x^2-q_y^2)}.
\end{equation}
Again, the largest contribution arises from $q_x\!=\!q_y\!=\!0$ and it is of the same size as for the nodal point (see upper panel of Fig.~\ref{fig:sigma_kq}). For $q_x\!=\!-\frac{\pi}{16}$ the first factor is largest at $q_y\!=\!0$, while the maximum of the second factor is located at $q_y\!=\!\pm q_x\!=\!\pm\frac{\pi}{16}$.

Since now {\sl both} factors depend quadratically on $\mathbf{q}$, the commensurate $q_y\!=\!0$ and the incommensurate $q_y\!=\!\pm\frac{\pi}{16}$ contribute similarly to the antinodal self-energy.

Hence, for the antinodal point incommensurate wave vectors near $\mathbf{q}=0$ (i.e., $\mathbf{q}\!=\!\boldsymbol{\pi}$ before the shift) are more relevant than for the node.
The lower panels of Figs.~\ref{fig:sigma_kq} and \ref{fig:sigma_kq_int} confirm this.
The calculation also explains why contributions from different $\mathbf{q}$ are symmetrical for the antinode, which is in general not the case for the nodal point (cf. both panels of Fig.~\ref{fig:sigma_kq}).

From a more physical perspective, our calculations show that the nodal-antinodal dichotomy originates in the van Hove singularity of the square lattice. Of course, the latter corresponds to a flat band structure at the antinodal point, which implies a vanishing first derivative. A more intuitive way to understand the calculation is therefore that the scattering from one antinodal point to another does not require a fine-tuning of the connecting (bosonic) momentum, because the associated single-particle energies vary only with the square of the deviation from the nesting vector $\boldsymbol{\pi}$. On the other hand, for generic Fermi vectors, including the nodal points, the single-particle energies vary linearly and hence scatterings do require fine-tuning.
See also Fig.~\ref{fig:sketch}.

%


\end{document}